# SCIENTOMETRICS AND SCIENCE STUDIES: FROM WORDS AND CO-WORDS TO INFORMATION AND PROBABILISTIC ENTROPY

Loet Leydesdorff

Department of Science and Technology Dynamics

University of Amsterdam

Nieuwe Achtergracht 166

1018 WV AMSTERDAM

The Netherlands

October 1995

# SCIENTOMETRICS AND SCIENCE STUDIES:

# FROM WORDS AND CO-WORDS

# TO INFORMATION AND PROBABILISTIC ENTROPY

The tension between qualitative theorizing and quantitative methods is pervasive in the social sciences, and poses a constant challenge to empirical research.  But in science studies as an interdisciplinary specialty, there are additional reasons why a more reflexive consciousness of the differences among the relevant disciplines is necessary.   How can qualitative insights from the history of ideas and the sociology of science be combined with the quantitative perspective?  By using the example of the lexical and semantic value of word occurrences, the issue of qualitatively different meanings of the same phenomena is discussed as a methodological problem.   Nine criteria for methods which are needed for the development of science studies as an integrated enterprise can then be specified.  *Information calculus* is suggested as a method which can comply with these criteria.

*Introduction*

The topical relations of science studies with issues in the philosophy of science make the clarification of the relation between theory and methodology urgent for the further development of *empirical* research in this interdisciplinary field.   In my opinion, the delineation of empirical science studies from older traditions of theory and philosophy of science can only be warranted if speculative reflection can be systematically supplanted by methodological exposition.

Furthermore, science studies develop in a science policy context so that analytical and programmatic questions can easily be confounded.   For example, in evaluation studies the primary question is typically whether the stimulation of a research programme has led to substantive developments, and not whether the additional funding has increasingly generated measurable activities.[21]   From this perspective science studies can afford to ignore philosophical questions about science and its progress only but at the price of becoming a rather trivial enterprise in itself.[11]   However, one needs methods for bringing together systematically the results from more normatively oriented types of analysis of science with those from more empirically oriented ones, *yet without confounding the two*.

The methodological reflection enables us to introduce the necessary distinctions in terms of domains and research questions.[13]   I shall argue in this communication that the discussion in scientometrics about words and their co-occurrences ("co-words") has provided us with a model for relating different types of theory with respect to the data.   This model for interdisciplinary science studies is then evaluated on its methodological implications.

*From "Words" to "Information"*

Word occurrences and their distributions are susceptible to statistical interpretation, *and* word occurrences in sentences have a semantic meaning.   Both the information content of the distribution and the meaning of the words may change over time, independently and/or in relation to each other.   Thus, we encounter the problem of relating different types of theory with respect to the same data in the study of words and their co-occurrences, at a very concrete level (e.g., Callon *et al.* [3 and 4]).

Already in 1955, Bar-Hillel [1] hinted at the possibility of a single *information calculus* enabling us to understand these two types of change in relation to each other.   The statistical interpretation takes the occurrence of a word as an instance of this word as a nominal variable. The occurrences and the patterns in their distributions inform us in a very basic sense about the text as a system of signals defined at the word level.[1]   Bar-Hillel, however, argued that the smallest unit of meaning is not the word itself but the sentence.   ("Half sentences often do not have meaning.")   Therefore, he wished to look at *words in sentences*. Words can have different meanings in sentences because of their different positions, argumentative functions, etc.

When sentences are taken as units of analysis, the unit of observation can still the word. But the relevance of an occurrence is differently evaluated, since the systems under study are different. The occurrence of a word in a sentence is no longer the instance of that word itself as a nominal variable, but the instance of another category which must be specified in terms of a theory of meaning of words in sentences. While in the former case we were interested in structure in textual data, this latter theory refers to structure in meaning among language users. Thus, the scheme by which an author like Bar-Hillel would rate word occurrences is different: two different words may be instances of the same variable in a scheme which assesses word occurrences in terms of meaning (e.g., synonyms), and the same word may be rated on different variables in two different instances (e.g., because of its position in the sentence).

---

[1] One could also have studied the system of signals at the character level (cf. Shannon [18]), but in science studies substantive reasons have been specified for looking at the aggregation process and the dynamics of networks in terms of words and their co-occurrences (e.g., [5]; [11]).

In summary, researchers using different theories can be expected to generate from the same data two (ore more) different relative frequency distributions. This is the crucial point for the methodological intervention: the same data, but multiple relative frequency distributions! Two analysts see something different in the same data. The data are made relevant to two or more histories.[2] However, in each case the result is eventually a relative frequency distribution: both analysts are able to ascertain whether something, the relevance of which can be specified in terms of their respective theories, is the case or not; or has occurred or not; or is to be expected or not.

Otherwise incompatible theories do not have to contradict one another in terms of their results. Theories guide us in collecting the data, and in providing us with an interpretation of the results. But the results of the measurement are comparable even if the guiding theories are mutually incommensurate.[cf. 8] The formal basis of the respective inferences offers us a common ground for a data theoretical comparison in terms of the quality of the representation.

---

[2] See for the concept of "multi-historicity" in relation to various reference systems: Luhmann [14], at pp. 88ff.
3

*What is indicated by the indicators?*

The relative frequency distributions are the results of (sometimes implicit) theoretical assumptions concerning the subject of study.  For example, a co-word mapping can be considered as a representation of a field of science.  As noted, the unit of analysis that is represented in the representation may be different from the representation itself.  In general, a representation provides us with a window on the represented system by using the representing system.  The various systems, however, can also be theoretically specified.  By paying both attention to the (methodological) quality of the representation and the relevant substances, a dual perspective on theory and methods can be developed that enables us, among other things, to bridge the gap between qualitative and quantitative science studies.

On the qualitative side of the field, only contributions which involve at least the lowest measurement scale, i.e. description in nominal categories, can be considered as part of *empirical* science studies.  The (e.g. historical) description provides us with a zero-order explanation: it indicates a possible explanation.  (Note that philosophical contributions can be relevant for empirical science studies by using this criterion.)  Whereas measurement in terms of categories is only a minimal requirement, more precise measurement is often possible.  For example, in an aggregate one may be able to specify not only whether something was the case, but also how often it was so.  The number of nominal instances can then be counted at the interval level.

On the quantitative side of science studies, the need for analysis in terms of structural units of science and in terms of various dimensions is usually recognized (see, e.g., [15]; [19]), but hardly ever are these latent units made subject to systematic theorizing. The emphasis is on the organization of data, using various methods of multi-variate analysis, and on graphic representations of the results (e.g., "mappings"). Statistical methods are often black-boxed in sophisticated computer programmes, which the researchers–in many cases legitimately–use to analyze their data. However, the choice of various parameters in statistical methods, e.g., similarity criteria and clustering algorithms, is not always discussed with reference to the theoretical questions, and thereby a vision of methods as only a kind of magical toolbox tends to be reinforced.[cf. 18]

Methods specify the procedures by which data can be related to theories. Thus, they provide us with means to reflect on the quality of the representation. Most methods, however, have been developed for theoretical purposes; they are based on specific assumptions. For example, co-citation analysis and co-word analysis use a *relational* algorithm like single linkage clustering.[3; 4; 19] A relational algorithm, however, cannot be used for indicating the structural dimensions of a network, since this requires an analysis in terms of *positions*.[2] Thus, the relational co-word and co-citation maps should not be considered as a tool for studying the structure of science.[9] In other words, the choice of a method implies a (sometimes implicit) hypothesis with respect to the data.

The variety in clustering algorithms and similarity criteria signals the variety of reconstructions that is possible on the basis of scientometric data. Each reconstruction refers to a theoretical appreciation, and therefore remains a hypothesis. In this respect the scientometric reconstruction has a similar status as a narrative reconstruction [cf. 6]; it differs mainly in the extent to which it can be systematically tested given a theoretical perspective. But as noted, the data can be interpreted in different theoretical contexts. For example, the scientific journal system by operating produces a yearly distribution of publications (citations, etc.) over nations. What do these distributions indicate? This scientometric data can be assessed with reference to the journal system, the international division of scientific labour, and/or the development of the international science system. With reference to which system(s) do the indicators exhibit change, and which systems were (sometimes implicitly) assumed to remain stable during the period under study?   Or, in terms of the example of the previous section, does a change in co-word structure reflect a change in language usage or does it reflect a change in the conceptual apparatus that is indicated by these co-words?

The observed distributions inform the reflexive analyst with reference to the *hypothesized* systems under study.   The hypothesized systems appear in the reconstruction as the grouping variables, the factor designations or the cluster structure that the analyst attributes (or instructs the computer to attribute) to the data. One does not study aggregated journal-journal citations as data, but one analyzes them as distributions *in terms of*, for example, specialties.   Accordingly, one is not able to observe the specialties in the data without an analytical assumption, i.e., an expectation that the assumed existence of specialties may induce a structure in the data.

In other words: the data exhibit the mutual information or the co-variation between what is being grouped and the grouping variable, or between the represented system and the representing system.[3]   For example, a body of scientific literature can be analyzed in terms of words and co-words.   The occurrences indicate both the documents under study, and the vocabulary used in these documents.   The observable occurrences can be considered as events that inform us about a co-variation between the document set and the available vocabulary. However, information about the co-variation *plus* information about the remaining variation in each of the co-varying systems is needed for improving our prediction about the respective system's future behaviour.   The remaining variation is not visible in the event, but it can be specified as an expectation by the analyst, either directly or mediated through the choice for a computer program.   The observed distributions enable us to test the quality of the expectations given these specifications.   Thus, the measurement informs us, if we are able to appreciate the information in theoretical terms.

---

[3] Although differently defined the mutual information and the co-variance are both measures of the uncertainty in the co-variation.

*Methodological implications*

A model has to show its strength in the prediction. The computer, however, allows us to test sets of models by using, for example, different algorithms in analyzing the data. Since all the models remain informed conjectures, one may wish to develop methods which enable us to compare not only results with expectations given one model or another, but also among different representations. While the various reconstructions are based on different assumptions, the data analyst may wish to uncouple as far as possible the heuristic function of theory in data collection and its appreciative function in the interpretation of empirical results, on the on hand, and the formal analysis, on the other hand. Is it possible to develop second-order methods which are content-free with reference to the reconstructions under study?

Note that the development of such methods would not only serve the comparison with hindsight. Content-free methods might have functions also in the empirical research process. A researcher often develops his/her research question when actually performing the empirical part of the research. Reflection on preliminary results easily leads to a reformulation of the research question. The parameters which were appropriately chosen for one specific research question, however, will almost never be the most appropriate choice for the next case, especially when one not only addresses a different domain, but also different variables, or when one extends the analysis to other levels of aggregation (i.e., to use different grouping rules). As noted, these parameters often refer to theoretical assumptions, and thus, the question may rise again of how to compare among results based on different assumptions.

For example, the most appropriate or at least, the currently best available technique for analyzing the relations among actors, is in terms of various types of network analysis. By doing so, one is able to compute the densities of the networks, centralities, and structural equivalencies in the positions of the various actors involved.[e.g., 2] However, as soon as one addresses the question of how scientific journals relate to each other, it becomes problematic, or at least metaphorical, to think of journals as actors who cite one another, and to analyze aggregated citation networks among them in terms of the *same* measures of centrality, density, and structural equivalence.[24]   For example, the question can be raised how to handle diagonal values in journal-journal citation matrices (are these "self-citations"?); but this difficulty indicates the underlying conceptual problem of the difference between self-citation on the part of individuals and of (articles within) journals (see, e.g., [16]).   These methodological issues indicate conceptual problems.[12]

One strategy to follow is to use only one type of analysis for the various problems involved.   Actually, many researchers have an inclination to address a new problem with methods that are familiar from previous research.   However, the problems of different dimensions, different levels of aggregation, and heterogeneous units of analysis will predictably lead to a failure of this approach in more complex instances. For example, if the researcher chooses one type of multi-variate statistics for the scientometric mapping of data-matrices composed from large document sets, s/he may thereafter wish to know whether one can decompose the total set into subsets which more pronouncedly exhibit the previously retrieved structure, and others which are more random.   Which units contribute most to the structure?   However, while the number of citations or the number of co-occurrences of words (co-words) in a database is a frequency, and therefore may be analyzed as an interval scaled variable, at the level of each of the composing texts these same co-occurrences may be dichotomous: the citations, c.q. words, either occur or not.   How can one then correlate the results of the metric and non-metric analyses?   More generally: How can one compare reconstructions in which different types of analysis lead to the same results, and others in which this is not the case?   The intrinsic problems of statistics may distract us from the substantive research questions.

In summary, although one is able to specify the limitations of a given method, one is usually not able to specify *a priori* whether one would like to extend the analysis on the basis of the results of a first analysis to a domain which is beyond the reach of that method.   Any extension of the research may imply a shift in the level of aggregation, the measurement scale, and the relevant variables.

*Methodological requirements for interdisciplinary science studies*

The above considerations allow us reflexively to list some methodological requirements for science studies as an integrated enterprise. First, it follows from the considerations in the previous section, that methods should enable us to vary over levels of aggregation, measurement scale, and relevant variables. As noted, the specification of other relevant variables may imply the attribution of the same data to other possible units of analysis or the addition of other data with reference to the same unit of analysis. In the latter case, methods should preferably allow for the specification of the increment in the information.

In the case of another unit of analysis, methods should allow us to perform secondary analysis by using previous data collections and data analysis. Only if this latter requirement can be warranted, one is able to build on the results of the many case studies performed for other (e.g., policy) reasons. This is an urgent question for scientometrics, indeed, since data collection is often too expensive for fundamental research, while contract money is sometimes available for indicator work.

Let me more systematically summarize the methodological requirements that were derived above for integrating theoretical work in science studies with the quantitative perspective provided by scientometric methods:

1.Methods should make it possible to actively import data and results (e.g., descriptions, facts, trendlines) from other types of studies. One can call this *the requirement of secondary analysis*: Data analysis should preferably support the translation among the various paradigms which are used in interdisciplinary science study.

2. Second-order methods should allow for variation in the types of theories and methods which use the same or similar data. They should therefore be reflexive with respect to the research process, and not prescriptive in any strong sense. In particular, methods should allow for the appreciation of qualitative descriptions. This may be called the *requirement of multiple paradigms*. (Of course, if one wishes to use a particular method one should use it correctly.)

In addition to these two requirements, one can also specify:

3. *the requirement of aggregation and decomposition.* Methods should allow us to control for the relations among levels of aggregation.[cf. 23]

This latter requirement, however, holds not only when we move among levels of aggregation in one dimension. In empirical science studies, the researcher may wish to import, for example, information about developments in literary structures at the field level (e.g., journal structures) into a research design which focusses on social processes at the level of research groups. The units of analysis at the different levels of aggregation are then heterogeneous. Thus, this leads to a fourth requirement which is a composite of the above requirements, i.e.:

4.     *the requirement of "heterogeneous nesting"* (cf. [3]).

As noted, we do not *a priori* require measurement to be more precise than nominal, since we wish to allow for historical and explorative research. However, one would like to be able to use any information that can be achieved by more accurate measurement. (For example, in order to address questions like whether the described developments are *significantly* different from comparable ones.) Therefore, in addition to the above specified requirements, we may now specify as a fifth criterion for more integrative methods in science studies a permissive requirement with respect to the measurement technique:

5. Methods should allow for variation in the measurement scale of observations, but save any additional information from better measurement. This *requirement of neutrality in terms of the measurement scale* asks technically for a *non-parametric method*.

Actually, the use of non-parametric statistics is convenient for the import of scientometric data since the distribution of this data is often skewed. Most multi-variate statistics, however, is based on assumptions concerning normality in the distributions.[4] Therefore if one relaxes this assumption, one has additionally to specify the following requirements for the methods that one looks for:

6. *requirement of multi-variate statistics*, i.e., methods for science studies should allow us to develop non-parametric equivalents of clustering algorithms, etc., on datasets which we can also compose and disaggregate. Higher level results should be interpretable in terms of lower level results, and *vice versa*.

In science studies we are interested not only in these complex data structures at each moment in time, but also in their development over time. Therefore, in addition to providing us with a full equivalent of "multi-variate analysis," methods should provide us with possibilities for studying time series of data, to make predictions, and to reconstruct. This leads to the formulation of the following two further requirements:

7. *the requirement of dynamic analysis*, i.e., methods should allow us not only to analyze (multi-variate) data in slices at each moment in time, but systematically to account for change in the various dimensions, and in relation to overall development.

8. *the requirement of reconstruction*, i.e., methods should enable us not only to analyze dynamically and multi-variately, but also to investigate irreversibilities in the data (e.g., path-dependent transitions). Note that the formulation of this requirement is in itself neutral to the question of whether one analyzes historical descriptions or scientometric data sets.

---

[4] Many scientometricians treat their data as if they were normally distributed. For example, in evaluation studies the mean is often used as a norm.

Finally, with respect to the data we may formulate one additional criterion which pertains to the specificity of the domain of science studies:

9. *the requirement of virtually no systems limitations on the number of variables*, since methods should allow us to study complex phenomena and/or large communities and archives, i.e. many variables, at both aggregated and decomposed levels.

Methods for science studies should preferably not only meet one or a few of these requirements, but make it possible for the analyst to integrate results from studies in which only a subset of these requirements are needed. Therefore, methods should in principle comply with all these requirements. This means that one type of analysis is systematically relatable to another in terms of specifiable transformations.

As already conjectured by Bar-Hillel [1], Shannon's [18] information theory can be of help, indeed. The expected information content of a distribution or its probabilistic entropy is, among other things, non-parametrical, content-free, and definable in statical and dynamical measures. Theil [23] has extended Shannon's formulas to the multi-level and the multi-variate case (see also: [7]). Information theory seems to provide us with a useful method for complying with the specified criteria. The recent application of this method to a large set of problems in science studies makes it increasingly plausible that the development of science studies as a specialty can become a more integrated enterprise.[11]

*Summary and Conclusions*

The bridging of the gap between qualitative theorizing and the use of scientometric methods is only one among a set of requirements for the further integration of science studies.[13] On the one hand, both in qualitative and in quantitative research, the researcher has to specify categories (variables), levels of aggregation, and relevant time horizons, before any type of change can be described, or tested against the data. On the other hand, if the same data can reveal different dynamics, how are we to analyze these dynamics, both independently and in relation to one another?

By elaborating the example of the measurement of "meaning" in terms of word occurrences in the semantic tradition and the measurement of word distributions in the semiotic tradition,[3] I specified how to relate the different meanings of data in science studies. A set of criteria could be derived for methods in science studies which aim at integration, despite the noted differences at the theoretical and the methodological levels. By further reflection on some methodological issues, and issues in relation to the type of data involved, additional criteria for this purpose could be specified.

The use of probabilistic entropy as an integrative measure in scientometrics refers to theories about dissipation in potentially self-organizing systems (e.g., [20]; [22]). Scientometric indicators provide us with a rich domain in terms of complex and longitudinal data for testing hypotheses concerning cultural evolution.[25] The elaboration of this issue, however, reaches beyond the scope of this communication.[10]

*References*

[1] Bar-Hillel, Y. (1955). An Examination of Information Theory. *Phil. Sci.* 22, 86-105.

[2] Burt, R. S. (1982). *Toward a Structural Theory of Action*. New York, etc.: Academic Press,

[3] Callon, M., J. Law, and A. Rip, Eds. (1986). *Mapping the Dynamics of Science and Technology*. London: Macmillan.

[4] Callon, M., J.-P. Courtial, H. Penan (1993). *La Scientométrie*.   Paris: Presses Universitaires de France.

[5] Hesse, M. (1980). *Revolutions and Reconstructions in the Philosophy of Science*. London: Harvester Press.

[6] Hinton, G., J. L. McClelland, and D. E. Rumelhart, Distributed Representations. In: [27], 77-109.

[7] Krippendorff, K. (1986). *Information Theory. Structural Models for Qualitative Data*. Beverly Hills, etc.: Sage.

[8] Kuhn, T. S. (1962/1970). *The Structure of Scientific Revolutions*. Chicago: Chicago University Press.

[9] Leydesdorff, L. (1992). A Validation Study of LEXIMAPPE. *Scientometrics* 25, 295-312.

[10] Leydesdorff L. (1994). The Evolution of Communication Systems. *Int. J. Systems Research and Information Science* 6, 219-30.

[11] Leydesdorff, L. (1995).   *The Challenge of Scientometrics: the development, measurement, and self-organization of scientific communications.* Leiden: DSWO Press, Leiden University.

[12] Leydesdorff, L., and O. Amsterdamska (1990). Dimensions of Citation Analysis. *Science, Technology and Human Values* 15, 305-335.

[13] Leydesdorff, L., J. Irvine. and A.F.J. van Raan, Eds. (1989). The Relations Between Qualitative Theory and Scientometric Methods in Science and Technology Studies. Theme Issue of *Scientometrics* 15, 333-631.

[14] Luhmann, N. (1990). *Die Wissenschaft der Gesellschaft*. Frankfurt a.M.: Suhrkamp.

[15] Mullins, N., W. Snizek, and K. Oehler (1988). The Structural Analysis of a Scientific Paper. In: A. F. J. Van Raan (Ed.), *Handbook of Quantitative Studies in Science and Technology*. Amsterdam: Elsevier.

[16] Price, D. de Solla (1981). The Analysis of Square Matrices of Scientometric Transactions. *Scientometrics* 3, 55-63.

[17] Rumelhart, D. E., J. L. McClelland, and the PDP Research Group (1986). *Parallel Distributed Processing* Vol. I. Cambridge, MA/ London: MIT.

[18] Shannon, C. E. (1948). A Mathematical Theory of Communication," *Bell System Technical Journal* 27, 379-423 and 623-56.

[19] Small, H., E. Sweeney, and E. Greenlee (1985). Clustering the Science Citation Index Using Co-Citations II. Mapping Science. *Scientometrics* 8, 321-40.

[20] Smolensky, P. (1986). Information Processing in Dynamical Systems: Foundation of Harmony Theory. In: [27], 194-281.

[21] Spiegel-Rösing, I. (1973). *Wissenschaftsentwicklung und Wissenschaftssteuerung*. Frankfurt a.M.: Athenäum Verlag.

[22] Swenson, R. (1989). Emergent Attractors and the Law of Maximum Entropy Production: Foundations to a Theory of General Evolution. *Systems Research* 6, 187-97.

[23] Theil, H. (1972). *Statistical Decomposition Analysis*. Amsterdam-London: North Holland.

[24] Todorov, R., and W. Glänzel (1988). Journal citation measures: a concise review. *Journal of Information Science* 4, 47-56.

[25] Van Raan, A. F. J. (1991). Fractal Geometry of Information Space as Represented by Co-Citation Clustering. *Scientometrics* 20, 439-49.